\documentclass[aps,superscriptaddress,showpacs,nofootinbib,twocolumn]{revtex4}
\usepackage{bm}
\usepackage{xcolor}
\usepackage{graphicx,color}
\usepackage{amsmath}
\usepackage{amssymb}

\newcommand{\be}{\begin{equation}}
\newcommand{\ee}{\end{equation}}
\newcommand{\bea}{\begin{eqnarray}}
\newcommand{\eea}{\end{eqnarray}}

\begin{document}

%--------------------------------------------------------------------------------
\title{Information production in homogeneous isotropic turbulence}
%--------------------------------------------------------------------------------
 
\author{Arjun Berera}
 \email{ab@ph.ed.ac.uk}%Lines break automatically or can be forced with \\
\author{Daniel Clark}%
 \email{daniel-clark@ed.ac.uk}
\affiliation{%
 School of Physics and Astronomy, University of Edinburgh, JCMB,
\\King’s Buildings, Peter Guthrie Tait Road EH9 3FD, Edinburgh, United Kingdom.
}%

\date{\today}% It is always \today, today,

\begin{abstract}
 We study the Reynolds number scaling of the Kolmogorov-Sinai entropy and attractor dimension for three dimensional homogeneous isotropic turbulence through the use of direct numerical simulation. To do so, we obtain Lyapunov spectra for a range of different Reynolds numbers by following the divergence of a large number of orthogonal fluid trajectories. We find that the attractor dimension grows with the Reynolds number as Re$^{2.35}$ with this exponent being larger than predicted by either dimensional arguments or intermittency models. The distribution of Lyapunov exponents is found to be finite around $\lambda \approx 0$ contrary to a possible divergence suggested by Ruelle. The relevance of the Kolmogorov-Sinai entropy and Lyapunov spectra in comparing complex physical systems is discussed.
\end{abstract}

\maketitle

%\tableofcontents

One of the most striking features of turbulent fluid flows is their seemingly random and unpredictable nature. However, since such fluids are described by the Navier-Stokes equations, which are entirely deterministic in nature, their motion cannot be truly random. In reality, turbulent flows exhibit what is commonly referred to as deterministic chaos \cite{bohr2005dynamical, ott2002chaos}. The time evolution of such systems is characterized by an extreme sensitivity to initial conditions which has profound consequences for their predictability. 

Studying turbulence through the lens of chaos theory and the related, but wider encompassing, area of dynamical systems theory has its roots in the seminal work of Ruelle and Takens \cite{ruelle1971nature} alongside that of Lorenz \cite{lorenz1963deterministic}. This approach differs from the more standard statistical approach \cite{batchelor1953theory} in the sense that, instead of considering averaged properties of flows, we consider the properties of individual trajectories in a suitably defined state space of the system. Through such methods, a diverse range of problems in fluid dynamics have been studied including in weather and atmospheric predictability \cite{lorenz1969predictability, leith1971atmospheric, leith1972predictability}, as well as for the solar wind and other magneto-hydrodynamic systems \cite{horton2001chaos, kurths1991lyapunov, winters2003chaos, ho2019chaotic}. 

A central theme for a large proportion of the literature investigating the chaotic properties of homogeneous isotropic turbulence (HIT) is the concept of the Lyapunov exponents. Put briefly, these exponents describe the rate of exponential stretching and contracting in the state space and are thus intimately related to the aforementioned sensitivity to initial conditions. For a given system, there exist as many Lyapunov exponents as degrees of freedom, which for real world Eulerian fluid turbulence is presumably infinite. To date the majority of such work, at least in the case of direct numerical simulation (DNS), has been concerned with the calculation of only the largest Lyapunov exponent \cite{berera2018chaotic, boffetta2017chaos}. It is, however, possible to compute multiple exponents, leading to a partial Lyapunov spectrum, to obtain a more in-depth understanding of the chaotic properties of the system. Moreover, if all positive exponents are calculated, the Kolmogorov-Sinai (KS) entropy \cite{sinaui1959concept, kolmogorov1958new}, which quantifies the rate of information production, can be estimated.

This paper presents a systematic study of the KS entropy for homogeneous isotropic fluid turbulence providing, to our knowledge, the most precise measurement to date. Using DNS to study the exact evolution of the Navier-Stokes equations, we test a Reynolds number scaling relation derived by Ruelle \cite{ruelle1982large}. As such, our results are only limited by the resolution of our simulations and the enormous computational cost of these measurements. Unfortunately, this restricts us to a limited range of Reynolds numbers and thus the applicability to fully turbulent flows is not yet conclusive. The results in this work also offer one of the first numerically rigorous tests of the various mathematical relations and conjectures in the literature associated with the KS entropy. Moreover, these results provide a benchmark for comparing against approximate numerical methods for computing the KS entropy. 

It is of additional interest to compare the entropy scaling presented in this paper to that seen in condensed matter \cite{mutschke1993kolmogorov} and quantum field theoretic systems \cite{bianchi2018linear,hallam2019lyapunov}. Indeed, if a similar scaling with a suitably defined control parameter is observed, then through Ornstein's isomorphism theorem \cite{ornstein1974isomorphism} there may exist connections at the level of information production between these seemingly disparate systems. This approach strips away the qualitative features that may distinguish different systems of complexity and reduces them all to a common set of quantitative measures that can be systematically cross compared. Such an approach could be beneficial in utilizing theoretical knowledge for one type of system to understand another with similar information theoretic content, and likewise for transfer of applications between such related systems. To develop such a program, a key step is to have good measurements of the Lyapunov spectra and KS entropy for a diverse set of complex systems.  HIT is one such benchmark system, and another purpose of this paper is to provide this measurement for it. For a detailed discussion of such ideas see \cite{gaspard1993noise}.

 Previous studies in this area have typically relied on employing methods to reduce the number of degrees of freedom, for example by using shell models \cite{yamada1987lyapunov, yamada1988lyapunov, grappin1986computation} or highly symmetric flows \cite{van2006periodic}. For the case of DNS there has, however, been far less work completed in this area, presumably due to the high computational cost.  Nonetheless, a small number of low resolution studies have been conducted, although these were focused primarily on estimating the dimension of the attractor for the flow \cite{grappin1991lyapunov, keefe1992dimension}. Even at the low Reynolds numbers achieved in these studies, the dimension of the attractor, and by extension the number of positive Lyapunov exponents, is in the order of hundreds. Hence, HIT is a distinctly high dimensional chaotic system and methods used for low dimensional systems are likely to give poor estimates of both the KS entropy and the attractor dimension \cite{eckmann1992fundamental}. 

Theoretical predictions for the KS entropy in HIT are typically mathematically complex. One of the most notable of these predictions was proposed by Ruelle \cite{ruelle1982large}, which can be simplified if we ignore the effects of intermittency. This assumes the energy dissipation takes its average value everywhere in the flow, and works in the framework of Kolmogorov's 1941 theory (K41) \cite{kolmogorov1941local}. In doing so, we come to the relation for the KS entropy, $h_{KS}$, in HIT \begin{equation}
h_{KS} \sim \frac{1}{\tau_{\eta}}\left(\frac{L}{\eta}\right)^3 \sim \frac{1}{T}Re^{\frac{11}{4}}.
\end{equation} Here, the Kolmogorov microscale is given by $\eta = (\nu^3 / \varepsilon)^\frac{3}{4}$ and the corresponding timescale by $\tau_{\eta} = (\nu /\varepsilon)^{\frac{1}{2}}$ in which $\nu$ and $\varepsilon$ are the viscosity and energy dissipation respectively. The integral length scale, $L = (3\pi/4E)\int \mathrm{d}k \, E(k)/k$, gives the size of the largest eddies in the flow \cite{batchelor1953theory} and $\text{Re} = UL/\nu$ is then the integral scale Reynolds number, where $U$ is the rms velocity. Finally, the large eddy turnover time is defined as $T = L/U$. 

Deviations from K41 have been well studied in the literature \cite{anselmet1984high, gotoh2002velocity, kaneda2003energy, ishihara2009study}, with both intermittency and finite Reynolds number effects claimed to be possible causes \cite{kolmogorov1962refinement, frisch1978simple, benzi1984multifractal, mccomb2014homogeneous}. In studies of the maximal Lyapunov exponent \cite{berera2018chaotic, boffetta2017chaos} for HIT, different scaling behavior than that predicted by both K41 \cite{ruelle1979microscopic} and the multi-fractal model \cite{crisanti1993predictability} was observed. As such, exact agreement with this simplified prediction for the KS entropy would be very unexpected, and indeed our results also display a conflicting scaling behavior.

In order to conduct a model-independent test of this prediction, we perform DNS of the incompressible Navier-Stokes equations, using a fully dealiased pseudospectral code \cite{yoffe2013investigation} in a periodic box of side length 2$\pi$, \begin{equation}
\begin{split}
\partial_t \boldsymbol{u} + \boldsymbol{u} \cdot \boldsymbol{\nabla} \boldsymbol{u} &= -\boldsymbol{\nabla}P + \nu \nabla^2 \boldsymbol{u} + \boldsymbol{f}, \\
\boldsymbol{\nabla} \cdot \boldsymbol{u} &= 0.
\end{split}
\end{equation} Here $\boldsymbol{u}$ and $P$ are the fluid velocity and pressure fields  respectively and $\boldsymbol{f}$ is an external forcing defined as \begin{equation}
\boldsymbol{f}(\boldsymbol{k},t) =\begin{cases}
 (\varepsilon/2E_{f})\boldsymbol{u}(\boldsymbol{k},t) \quad &\text{if } 0 < k \leq k_f, \\
 0 \qquad &\text{else,}
\end{cases}
\end{equation} where $E_f = \int_{0}^{k_f}\mathrm{d}k \, E(k)$ is the energy in the forcing band. The main advantages of this form of forcing are that it allows for the energy dissipation, $\varepsilon$, to be set exactly and, since it simply feeds the velocity field back into itself, it does not introduce a stochastic element to the system. In all cases we have set $\varepsilon = 0.1$.

In this work we take the KS entropy to be given, as is standard \cite{ott2002chaos}, by the sum of positive Lyapunov exponents \begin{equation}
h_{KS} = \sum_{\lambda_i > 0} \lambda_i.
\end{equation} As such, we need to measure a number of exponents for each case. This leads to various difficulties: a priori we do not know ahead of time the number of positive exponents, the method used to obtain exponent values requires many iterations for averaging, and each exponent requires the simultaneous integration of another velocity field. Consequently, the computational cost involved in computing the KS entropy for fully resolved turbulent flows is high, and indeed we will show that even at moderate Re the number of positive exponents is in the thousands. 

We make use of the algorithm proposed by Benettin \cite{benettin1980lyapunov} in order to measure a large number of exponents, which we will briefly summarize here. After allowing time for the flow to reach steady state, we make $M$ copies, labelled $\boldsymbol{u_i}$, $i = 1 \dots M$ , of the reference velocity field, $\boldsymbol{u_0}$. Each copy is then perturbed using a Gaussian vector field with zero mean and a variance of size $\delta_0$, chosen such that the perturbation may be considered infinitesimal. To measure the exponents we use the finite time Lyapunov exponent (FTLE) method \cite{ott2002chaos} and track the growth of the difference fields $\boldsymbol{\delta_i}(t) = \boldsymbol{u_i} - \boldsymbol{u_0}$, rescalling the difference to its original size at time intervals of $\Delta t$ \begin{equation}
\boldsymbol{u_i}(\boldsymbol{k}, \Delta t) = \boldsymbol{u_0}(\boldsymbol{k}, \Delta t) + \frac{\boldsymbol{u_i}(\boldsymbol{k}, \Delta t) - \boldsymbol{u_0}(\boldsymbol{k}, \Delta t)}{\delta_0},
\end{equation} such that each perturbation continues to grow in the correct direction. The FTLEs are then given by \begin{equation}
\gamma_i(\Delta t) = \frac{1}{\Delta t}\ln \left(\frac{|\boldsymbol{\delta_i}(\Delta t)|}{\delta_0} \right),
\end{equation} and the Lyapunov exponents $\lambda_i$ are found by averaging over many iterations. As it stands, this algorithm will simply measure the largest Lyapunov exponent $M$ times, due to a tendency for all difference fields to align along the direction of the state space associated with this exponent. To prevent this, we make use of the modified Gram-Schmidt algorithm to orthogonalize the $\boldsymbol{\delta_i}$ after each measurement of the $\gamma_i$. By repeating this procedure for an infinite number of iterations, the exponents would become ordered such that $\lambda_1 > \lambda_2 > \dots > \lambda_M$. In our results, due to the finite number of iterations, the spectra are not monotonically decreasing, but reasonable ordering is achieved as in \cite{keefe1992dimension}. This can also be seen in figure \ref{fig2}. Due to this ordering property, we can ensure we have obtained all the positive exponents for the system by choosing $M$ such that a tail of negative exponents persist after averaging. The number of operations in the orthogonalization step scales with $M^2$ and thus quickly becomes computationally prohibitive. 

\begin{figure}
    \includegraphics[width=\columnwidth]{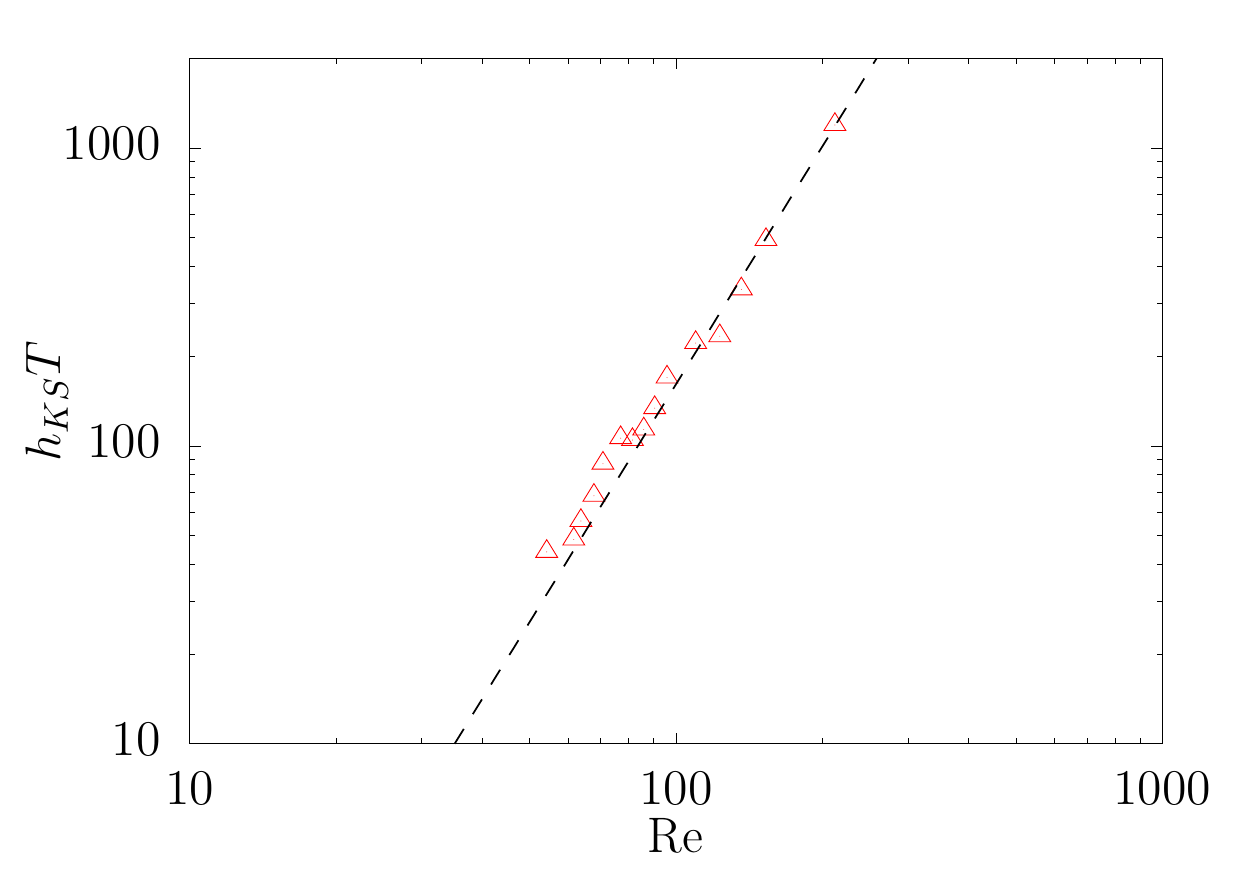}
    \caption{\label{fig1}The main plot shows $\text{Re}$ vs $h_{KS}T$ with the fit $0.0008\text{Re}^{2.65}$ as a dashed black line. To improve clarity, errors are not shown on this log-log scale.}
\end{figure}

Our results span a range from $\text{Re}\approx50$ to $\text{Re}\approx212$. 
Unfortunately, the number of positive exponents scales so quickly it would 
take excessive computational effort to explore higher Re flows.
To guarantee our data-sets are fully resolved, we maintain 
a minimum of $k_{\textrm{max}} \eta \geq 1.25$, where $k_{\textrm{max}}$ is 
the maximum wavenumber in the simulation determined by dealiasing using the two-thirds rule \cite{orszag1971elimination}. As such, runs were performed on grids 
ranging from $48^3$ to $150^3$ collocation points. In doing so, we can be 
confident that the small scales of the flow are adequately resolved. This is 
a non-trivial requirement for such a study, as it was observed 
in \cite{keefe1992dimension} that the number of positive exponents 
measured can be underestimated if the resolution used is too low. After non-dimensionalizing the KS entropy by multiplying by the large-eddy turnover time, we find our data is well fit by a power law of the 
form $h_{KS}T = C \text{Re}^{\alpha}$. A plot of this scaling behavior 
is shown in figure \ref{fig1}. The exponent is found to 
take the value $\alpha = 2.65 \pm 0.06$ with the constant given by $C= 0.0008 \pm 0.0002$. 

\begin{figure}
    \includegraphics[width=\columnwidth]{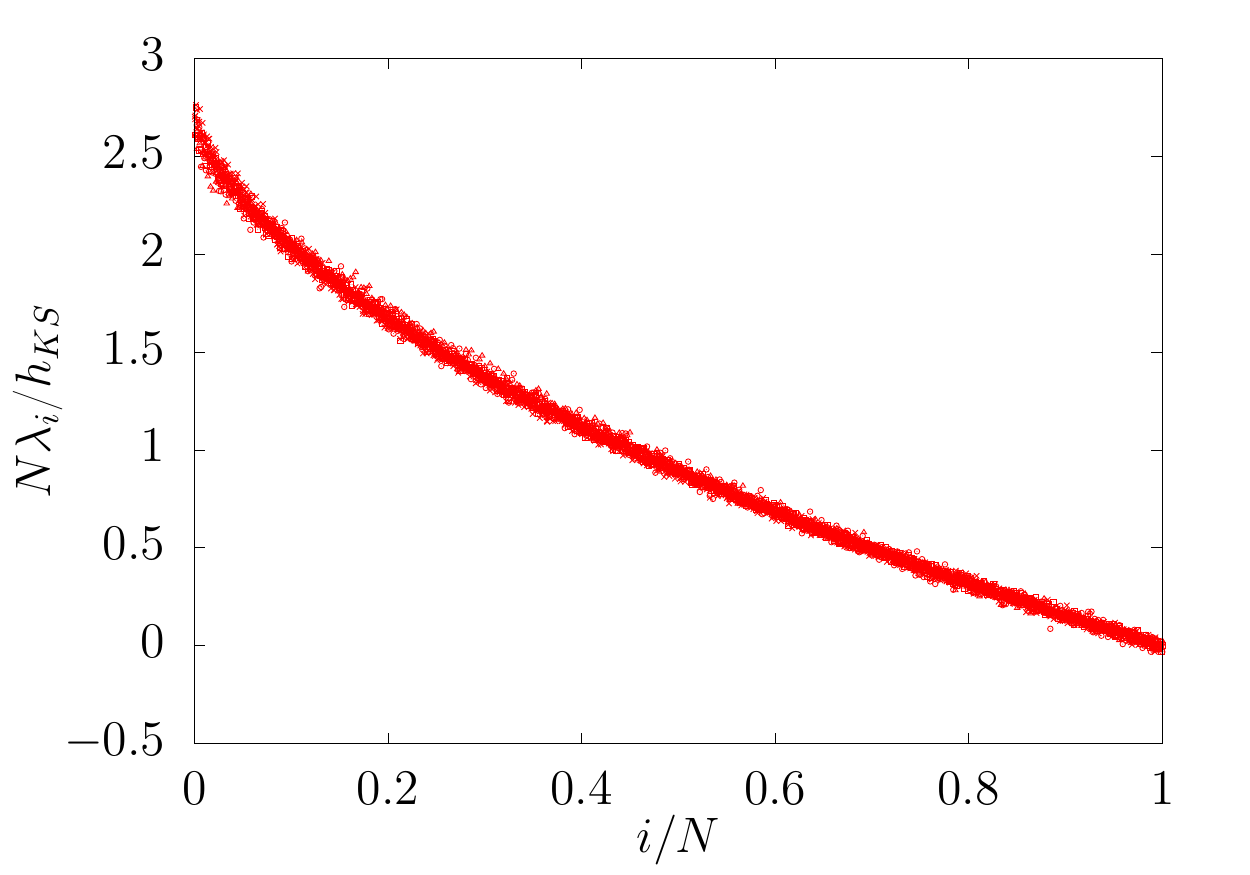}
    \caption{\label{fig2}Lyapunov spectra for $\text{Re} = 90 (\bigtriangleup)$, $123 (\circ)$, $153 (\Box)$ and $212 (\times)$ rescaled by $h_{KS}$ and the number of positive exponents, $N$. Using this normalization, Lyapunov spectra measured at different Re are shown to collapse onto the same curve, indicating a scaling property.}
\end{figure}

 Previous studies using shell models \cite{yamada1988lyapunov, grappin1986computation} found the Re scaling behavior for both the maximal Lyapunov exponent and the KS entropy to be the same and to follow the K41 prediction of $\text{Re}^{0.5}$ at odds with what is observed in our DNS. Meanwhile, using a multi-affine field \cite{boffetta2002predictability}, it is shown that the space-time entropy of the entire velocity field should scale according to the simplified prediction we are considering here. The reason for the large discrepancy between these results is well elucidated in \cite{wang1992varepsilon}, where on dimensional grounds they find the shell model result corresponds to measurement of the velocity at a single point, whilst the Ruelle result is for the KS entropy of the entire field. Since in our DNS we study the Lyapunov exponents for the entire velocity field, the Ruelle entropy result is the appropriate prediction for this work.
 
Our data shows the Re scaling exponent is less than expected when following the simplified prediction, although only slightly. Intermittency is often invoked to explain such differences, however, the multi-fractal model predicted deviations \cite{crisanti1993intermittency, aurell1996predictability} in the opposite direction than that observed in DNS \cite{berera2018chaotic, boffetta2017chaos} for the related largest Lyapunov exponent. As such, it is hard to make the claim that this is an adequate explanation for our results, although we cannot rule it out definitively. Another possible interpretation is that the discrepancy is due to finite Re effects. This is an attractive hypothesis, considering the modest Re values achieved, and indeed the K41 scaling is predicated on the existence of an inertial range, which will be limited in these simulations.  

Beyond the Re scaling of the KS entropy there are also conjectures related to the shape of the Lyapunov spectrum for fluid turbulence. The main question raised is related to the distribution of exponents about $\lambda \approx 0$, and whether or not it diverges around this point \cite{ruelle1983five}. In the $\beta$-model of turbulence, which attempts to extend the K41 theory to account for intermittency, the spectrum of exponents can become singular at $\lambda \approx 0$ \cite{ruelle1982large}. This simple fractal model has since been superseded by the multi-fractal model, but to our knowledge there does not exist a prediction for the distribution using this theory. Measurements of the spectrum in shell models seemed to confirm this divergence \cite{yamada1988lyapunov}, however, this was later suspected to be the result of the discretization in wavenumber space \cite{yamada1998asymptotic}, which is also present in the $\beta$-model.

\begin{figure}[]
    \includegraphics[width=\columnwidth]{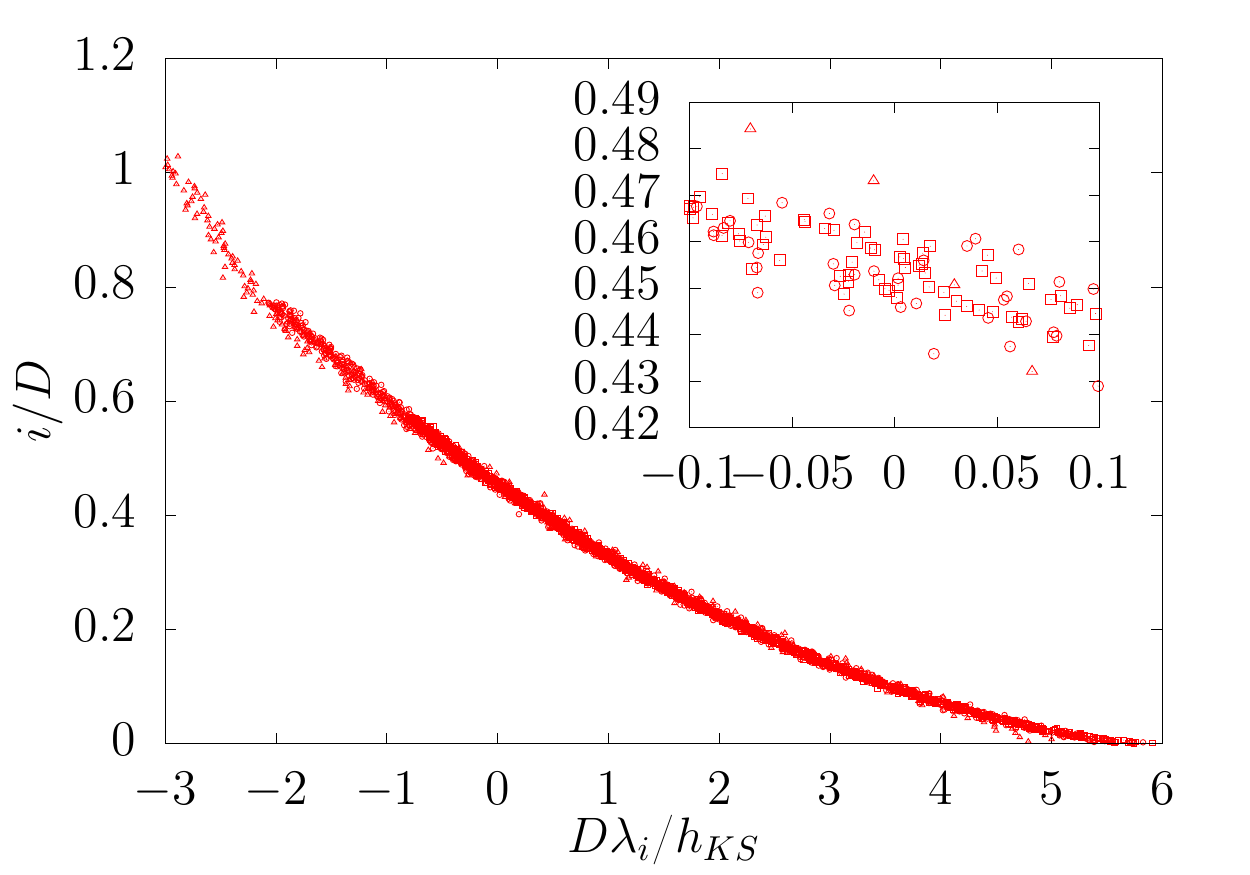}
    \caption{\label{fig3}Lyapunov spectra normalized by $h_{KS}$ and the (estimated) attractor dimension $D$. The inset shows the same plot zoomed in on the region around $\lambda \approx 0$, highlighting the lack of divergence. The spectrum represented by $\bigtriangleup$ is normalized using a directly computed dimension whilst the other cases use dimensions estimated from this.}
\end{figure}

For the range of Re tested in our DNS results there is no singularity in the Lyapunov spectra at $\lambda \approx 0$. To demonstrate this, we will need to obtain an estimate for the dimension of the attractor underlying the flow at each Re. Firstly, in order to make a sufficient estimate, we relate the attractor dimension to the Lyapunov exponents via the Kaplan-Yorke conjecture \cite{kaplan1979chaotic}. To do so, we require the integer $j$ such that \begin{equation}
\sum_{i=0}^{j} \lambda_i \geq 0, \quad \mbox{and} \quad \sum_{i=0}^{j+1} \lambda_i < 0,
\end{equation} and the attractor dimension is then \begin{equation}
D = j + \frac{\sum_{i=0}^{j} \lambda_i}{|\lambda_{j+1}|}.
\end{equation} Clearly, this requires far more exponents to be obtained than for the KS entropy calculation, meaning the computational cost becomes unfeasible at relatively low Re. Secondly, following \cite{keefe1992dimension}, we can identify an Re scaling behavior for the shape of the Lyapunov spectra in the region of positive exponents, which allows us to estimate the dimension of the higher Re cases. This can be seen in figure \ref{fig2} and if we consider the Re as a control parameter for the system, then this scaling is also seen in other systems of differential equations \cite{manneville1985liapounov, keefe1989properties}. Due to the spectra exhibiting a universal shape upon rescaling, we can take the attractor dimension to be proportional to the number of positive exponents. Explicitly, if we have one case where we know the attractor dimension, $D_1$, for a given Re value which has $N_1$ positive exponents, then for any other Re we can estimate the attractor dimension as $D_i \approx (N_i/N_1)D_1$, even if we only know $N_i$.

For a small subset of cases at lower Re, it is computationally feasible to measure enough exponents to make use of the Kaplan-Yorke dimension formula. Using these directly measured dimensions, we then estimate the attractor dimension for the higher Re cases. We find that in our $\text{Re} \approx 212$ simulation the attractor dimension is $D \approx 5704$. This result highlights that HIT is in a class of extremely high dimensional systems, even when compared to other chaotic fluid systems, such as Rayleigh-Bernard convection \cite{xu2016covariant}. Finally, in figure \ref{fig3} we present a different method of normalizing the spectra, this time using the estimated dimensions rather than the total number of positive exponents. This normalization was used in \cite{yamada1988lyapunov} to provide evidence of a divergence in the Lyapunov spectra around $\lambda \approx 0$ for a shell model. Therefore, as we set out to demonstrate for the case of the full evolution of the Navier-Stokes equations,
as studied in our DNS, figure \ref{fig3} shows no such divergence at this point across a range of different Re values.

\begin{figure}[]
    \includegraphics[width=\columnwidth]{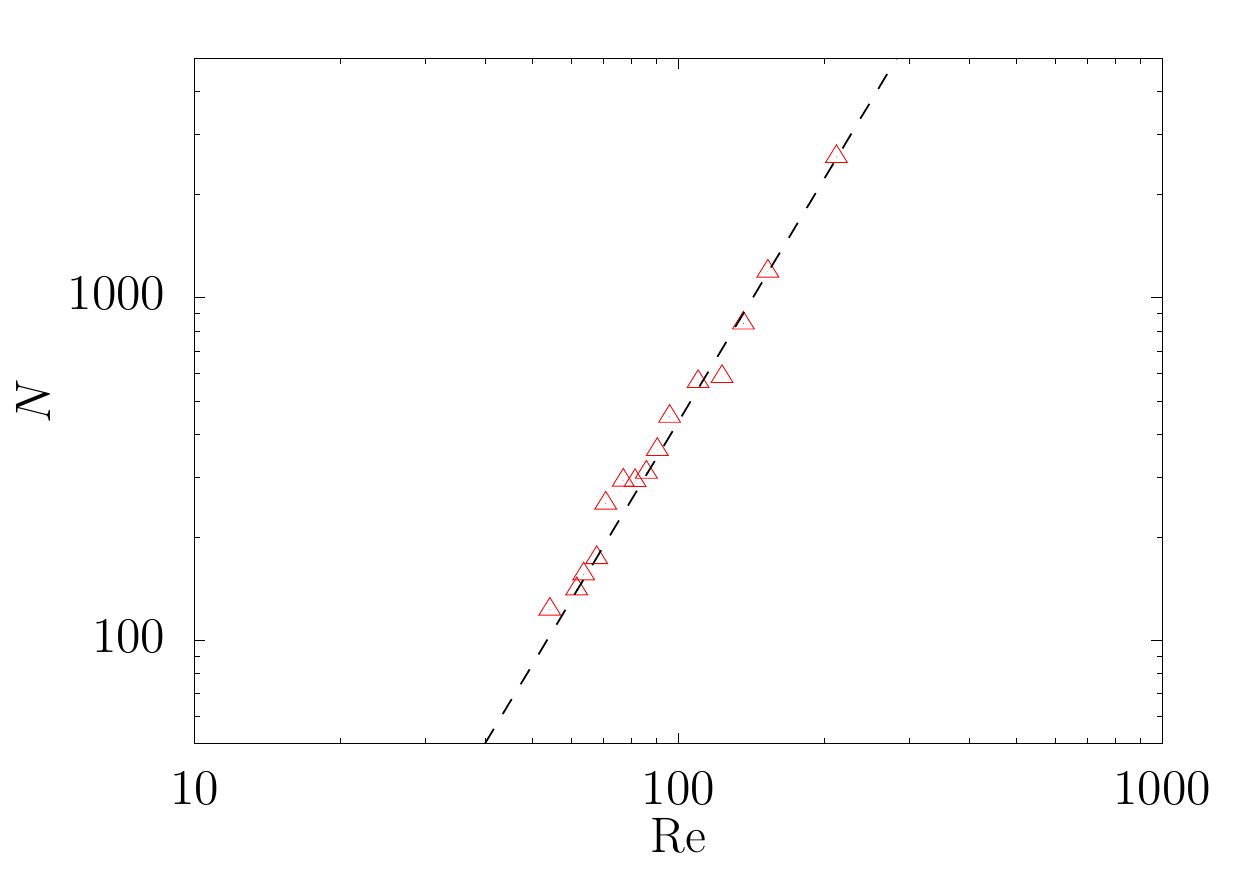}
    \caption{\label{fig4}Plot of Re against the number of positive exponents $N$. The dashed line shows the fit to the data of $0.008Re^{2.35}$}
\end{figure} 

We now turn to the scaling behavior for the number of positive exponents, $N$, with Re. Due to the shape similarity of the Lyapunov spectra, this scaling is also indicative of that for the attractor dimension, and hence the number of active degrees of freedom in the flow. Theoretically, the standard scaling for K41, based on purely dimensional grounds, was given by Landau and Lifshitz \cite{landau1959lifshitz} and gives $N \sim \text{Re}^{\frac{9}{4}}$. A number of additional mathematically derived estimates for the attractor dimension scaling have also been proposed \cite{constantin1985determining,gibbon1997attractor}, which predict scalings of $\text{Re}^3$ and $\text{Re}^{\frac{18}{5}}$ respectively. There also exists a prediction which makes use of the multi-fractal model for intermittency, where the scaling exponent is found to be slightly less than the K41 value \cite{meneveau1989attractor}. Remarkably, the estimate made on dimensional grounds is closest to what is observed in our DNS data, where we find the best fit to be of the form $N = b\text{Re}^{\gamma}$ with $\gamma = 2.35\pm0.05$ and $b = 0.008\pm0.002$. Thus, our data suggests that the number of positive exponents, and therefore the attractor dimension, grows with Re at a rate slightly faster than the Landau estimate. This result is in line with the findings in \cite{berera2018chaotic, boffetta2017chaos} concerning the largest Lyapunov exponent, whereby the correction to the K41 result is in the opposite direction when compared to the multi-fractal model value. As such, this suggests that if intermittency corrections are responsible for the deviations for K41 scaling seen here, then a different intermittency model, which captures the behavior seen in our results, is needed. 

If we consider that the number of operations in a DNS grows as Re$^{3}$ then the scaling of $N$ suggests that, as a lower bound, upon doubling Re, $2^{5.35} \approx 40$ times more operations are needed to measure the entropy. Thus, at present, the results in this study are on the limit of what is computationally possible.

To summarize, we have shown that in HIT the KS entropy exhibits a scaling law of the form $h_{KS}T \sim \text{Re}^{2.65}$ and this exponent is less than that predicted using the K41 theory. Further to this, Ruelle raised a question \cite{ruelle1982large} concerning the distribution of Lyapunov exponents in fluid turbulence and a possible divergence around $\lambda \approx 0$. Our results demonstrate that the Lyapunov spectrum for the incompressible periodic Navier-Stokes equations at the Re studied do not exhibit a divergence at this point. Moreover, we identify a Re independent shape of the spectra, allowing us to conjecture that there is no divergence at any Re. Additionally, we have also investigated the scaling of the number of positive Lyapunov exponents with Re. This, coupled with the Re independent spectra shape, allows us to also estimate the scaling behavior of the attractor dimension. We found that this quantity scales faster than is predicted using K41 physical arguments, and the opposite of what is predicted if intermittency is accounted for via the multi-fractal model. Caution should however be applied given the low Re values that could be studied via this method using current computing power. The nature of these results may exhibit differences to flows at higher Re which will be of interest to study when the requisite computational resources become available.

Through the use of numerical experiment, this study extends the understanding of the links between Eulerian turbulence and deterministic chaos. By utilizing the KS entropy as a measure of information production in HIT, it may also be possible to make a connection from HIT to information theory and related areas such as algorithmic complexity \cite{brudno1982entropy}, which we will examine in future investigations. Further to this, through the work of Ornstein \cite{ornstein1974isomorphism}, all systems satisfying certain technical conditions with the same KS entropy are isomorphic to each other. Following this line of argument, it may then be possible that other physical systems and HIT are in fact related at the level of information production. The results presented in this paper may then be relevant to a wide range of chaotic systems. In particular, the idea that other strongly coupled systems may be connected to HIT is already being explored through the ADS/CFT fluid-gravity correspondence \cite{rangamani2009gravity}, which may have a counterpart in information theory.

\begin{acknowledgments}
This work used the Cirrus UK National Tier-2 HPC Service at EPCC \cite{cirrus} funded
by the University of Edinburgh and EPSRC (EP/P020267/1).
A.B. acknowledges funding from the U.K. Science and Technology Facilities Council.
D.C. is supported by the University of Edinburgh.
\end{acknowledgments}


\begin{thebibliography}{99} 
 
\bibitem{bohr2005dynamical}
T. Bohr, M. H. Jensen, G. Paladin, and A. Vulpiani,
\textit{Dynamical systems approach to turbulence} (Cambridge University Press, 2005).

\bibitem{ott2002chaos}
E. Ott, \textit{Chaos in dynamical systems} (Cambridge University Press, 2002).

\bibitem{ruelle1971nature}
D. Ruelle and F. Takens, Commun. Math. Phys. {\bf 20}, 167 (1971).

\bibitem{lorenz1963deterministic}
E. N. Lorenz, J. Atmos. Sci. {\bf 20}, 130 (1963).

\bibitem{batchelor1953theory}
G. K. Batchelor, \textit{The theory of homogeneous turbulence} (Cambridge University Press, 1953). 
 
\bibitem{lorenz1969predictability}
E. N. Lorenz, Tellus {\bf 21}, 289 (1969).

\bibitem{leith1971atmospheric}
C. E. Leith, J. Atmos. Sci. {\bf 28}, 145 (1971).

\bibitem{leith1972predictability}
C. E. Leith and R. H. Kraichnan, J. Atmos. Sci. {\bf 29}, 1041 (1972).

\bibitem{horton2001chaos}
W. Horton, R. S. Weigel, and J. C. Sprott, Phys. Plasmas {\bf 8}, 2946 (2001).

\bibitem{kurths1991lyapunov}
J. Kurths and A. Brandenburg, Phys. Rev. A {\bf 44}, R3427 (1991).

\bibitem{winters2003chaos} 
W. F. Winters, S. A. Balbus, and J. F. Hawley, Mon. Not. R. Astron. Soc. {\bf 340}, 519 (2003).

\bibitem{ho2019chaotic}
R. D. J. G. Ho, A. Berera, and D. Clark, Phys. Plasmas {\bf 26}, 042303 (2019).

\bibitem{berera2018chaotic} 
A. Berera and R. D. J. G. Ho, Phys. Rev. Lett. {\bf 120}, 024101 (2018).

\bibitem{boffetta2017chaos}
G. Boffetta and S. Musacchio, Phys. Rev. Lett. {\bf 119}, 054102 (2017).

\bibitem{sinaui1959concept}
Y. Sinai, in Dokl. Akad. Nauk SSSR, Vol. {\bf 124} (1959) pp. 768–771.

\bibitem{kolmogorov1958new}
A. N. Kolmogorov, in Dokl. Akad. Nauk SSSR, Vol. {\bf 119} (1958) pp. 861–864.

\bibitem{ruelle1982large}
D. Ruelle, Commun. Math. Phys. {\bf 87}, 287 (1982).

\bibitem{mutschke1993kolmogorov}
G. Mutschke and U. Bahr, Physica D {\bf 69}, 302 (1993).

\bibitem{bianchi2018linear} 
E. Bianchi, L. Hackl, and N. Yokomizo, J. of High Energy Phys. {\bf 2018}, 25 (2018).

\bibitem{hallam2019lyapunov}
A. Hallam, J. Morley, and A. Green, Nature Comm. {\bf 10}, 2708 (2019).

\bibitem{ornstein1974isomorphism}
D. S. Ornstein, Yale University Press (1974).

\bibitem{gaspard1993noise}
P. Gaspard and X. J. Wang, Phys. Rep. {\bf 235}, 291--343 (1993).

\bibitem{yamada1987lyapunov} 
M. Yamada and K. Ohkitani, J. Phys. Soc. Jpn. {\bf 56}, 4210 (1987).

\bibitem{yamada1988lyapunov}
M. Yamada and K. Ohkitani, Phys. Rev. Lett. {\bf 60}, 983 (1988).

\bibitem{grappin1986computation}
R. Grappin, J. Leorat, and A. Pouquet, J. Phys. (Paris) {\bf 47}, 1127 (1986).

\bibitem{van2006periodic}
L. Van Veen, S. Kida, and G. Kawahara, Fluid Dyn. Res. {\bf 38}, 19 (2006).

\bibitem{grappin1991lyapunov}
R. Grappin and J. Leorat, J. Fluid Mech. {\bf 222}, 61 (1991).
 
\bibitem{keefe1992dimension}
L. Keefe, P. Moin, and J. Kim, J. Fluid Mech. {\bf 242}, 1 (1992).

\bibitem{eckmann1992fundamental}
J. Eckmann and D. Ruelle, Physica D {\bf 56}, 185 (1992).

\bibitem{kolmogorov1941local}
A. N. Kolmogorov, Cr Acad. Sci. URSS {\bf 30}, 301 (1941).

\bibitem{anselmet1984high}
F. Anselmet, Y. L. Gagne, E. J. Hopfinger, and R. A. Antonia, J. Fluid Mech. {\bf 140}, 63 (1984).

\bibitem{gotoh2002velocity}
T. Gotoh, D. Fukayama, and T. Nakano, Phys. Fluids {\bf 14}, 1065 (2002).

\bibitem{kaneda2003energy}
Y. Kaneda, T. Ishihara, M. Yokokawa, K. Itakura, and A. Uno, Phys. Fluids {\bf 15}, L21 (2003).

\bibitem{ishihara2009study}
T. Ishihara, T. Gotoh, and Y. Kaneda, Annu. Rev. Fluid Mech. {\bf 41}, 165 (2009).

\bibitem{kolmogorov1962refinement}
A. N. Kolmogorov, J. Fluid Mech. {\bf 13}, 82 (1962).

\bibitem{frisch1978simple}
U. Frisch, P.-L. Sulem, and M. Nelkin, J. Fluid Mech. {\bf 87}, 719 (1978).

\bibitem{benzi1984multifractal}
R. Benzi, G. Paladin, G. Parisi, and A. Vulpiani, J. Phys. A {\bf 17}, 3521 (1984).

\bibitem{mccomb2014homogeneous}
W. D. McComb, \textit{Homogeneous, isotropic turbulence: phenomenology, renormalization and statistical closures, Vol. 162} (OUP Oxford, 2014).

\bibitem{ruelle1979microscopic}
D. Ruelle, Phys. Lett. A {\bf 72}, 81 (1979).

\bibitem{crisanti1993predictability}
A. Crisanti, M. H. Jensen, G. Paladin, and A. Vulpiani, J. Phys. A {\bf 26}, 6943 (1993).

\bibitem{yoffe2013investigation}
S. R. Yoffe, Ph.D. thesis, University of Edinburgh, 2012, arXiv:1306.3408.

\bibitem{benettin1980lyapunov}
G. Benettin, L. Galgani, A. Giorgilli, and J. Strelcyn, Meccanica {\bf 15}, 9 (1980).

\bibitem{orszag1971elimination}
S. A. Orszag, J. Atmos. Sci. {\bf 28}, 6 (1971).


\bibitem{boffetta2002predictability}
G. Boffetta, M. Cencini, M. Falcioni, and A. Vulpiani, Phys. Rep. {\bf 356}, 367 (2002).

\bibitem{wang1992varepsilon}
X. J. Wang and P. Gaspard, Phys. Rev. A {\bf 46}, R3000 (1992).

\bibitem{crisanti1993intermittency}
A. Crisanti, M. H. Jensen, A. Vulpiani, and G. Paladin, Phys. Rev. Lett. {\bf 70}, 166 (1993)

\bibitem{aurell1996predictability}
E. Aurell, G. Boffetta, A. Crisanti, G. Paladin, and A. Vulpiani, Phys. Rev. E {\bf 53}, 2337 (1996).

\bibitem{ruelle1983five}
D. Ruelle, Physica D {\bf 7}, 40 (1983).

\bibitem{yamada1998asymptotic}
M. Yamada and K. Ohkitani, Phys. Rev. E {\bf 57}, R6257 (1998).

\bibitem{kaplan1979chaotic}
J. L. Kaplan and J. A. Yorke, in \textit{Functional Differential equations and approximation of fixed points} (Springer, 1979) pp. 204–227.

\bibitem{manneville1985liapounov}
P. Manneville, in \textit{Macroscopic Modelling of Turbulent Flows} (Springer, 1985) pp. 319–326.

\bibitem{keefe1989properties}
L. Keefe, Phys. Lett. A {\bf 140}, 317 (1989).

\bibitem{xu2016covariant}
M. Xu and M. R. Paul, Phys. Rev. E {\bf 93}, 062208 (2016).

\bibitem{landau1959lifshitz}
L. D. Landau and E. M. Lifshitz, Course of Theoretical Physics {\bf 6} (1959).

\bibitem{constantin1985determining}
P. Constantin, C. Foias, O. P. Manley, and R. Temam, J. Fluid Mech. {\bf 150}, 427 (1985).

\bibitem{gibbon1997attractor}
J. D. Gibbon and E. S. Titi, Nonlinearity {\bf 10}, 109 (1997).

\bibitem{meneveau1989attractor}
C. Meneveau and M. Nelkin, Phys. Rev. A {\bf 39}, 3732 (1989)

\bibitem{brudno1982entropy}
A. A. Brudno, Tr. Mosk. Mat. Obs. {\bf 44}, 124 (1982).

\bibitem{rangamani2009gravity}
M. Rangamani, Classical and quantum gravity {\bf 26}, 224003 (2009).

\bibitem{cirrus}
Cirrus, http://www.cirrus.ac.uk.


\end{thebibliography}
\end{document}